\documentclass[twocolumn,showpacs,preprintnumbers,amsmath,amssymb]{revtex4}

\usepackage{epsfig}
\def\be{\begin{equation}}

\def\ee{\end{equation}}

\def\lsim{\raise0.3ex\hbox{$<$\kern-0.75em\raise-1.1ex\hbox{$\sim$}}}

\def\gsim{\raise0.3ex\hbox{$>$\kern-0.75em\raise-1.1ex\hbox{$\sim$}}}

\begin{document}

\title{Cluster Percolation and Critical Behaviour in Spin Models and SU(N) Gauge Theories}

\vskip1cm

\author{Santo Fortunato}

%\maketitle

\vskip0.7cm

\affiliation{Fakult\"at f\"ur Physik, Universit\"at Bielefeld, D-33501 Bielefeld, Germany}

\vskip0.5cm

\begin{abstract}
\noindent

The critical behaviour of several spin models can be simply described
as percolation of some suitably defined clusters, or droplets:
the onset of the geometrical transition coincides with the critical point
and the percolation exponents are equal to the thermal exponents. 
It is still unknown whether, given a model, one can define
at all the droplets. In the cases where this is possible, the droplet definition
depends in general on the specific model at study and can be quite involved. 
We propose here a simple general definition for the droplets: they are 
clusters obtained by joining nearest-neighbour spins of the same sign 
with some bond probability $p_B$, which is the minimal probability that 
still allows the existence of a percolating cluster at the critical temperature $T_c$.
By means of lattice Monte Carlo simulations we find 
that this definition indeed satisfies the conditions required for the
droplets, for many classical spin models, discrete and continuous,
both in two and in three dimensions. In particular, our prescription
allows to describe exactly the confinement-deconfinement
transition of $SU(N)$ gauge theories as Polyakov loop percolation.

\end{abstract}



\maketitle

\vskip0.7cm

\section{Introduction}

The study of critical phenomena is one of the most
fascinating topics in physics: phase transitions are 
processes that physicists continuously encounter in their
investigations of nature, and they can occur in an incredible
variety of systems. 
A phase transition is basically a change in the order of a system: by varying
some parameter (usually the temperature), the microscopic constituents
of the system, that we here call `spins',  
choose a different way of staying together. If we go from 
high to low temperatures, we see that one passes from a situation
in which the behaviour of each spin is totally independent of the others, to
a situation in which several spins are correlated with each other and form
ordered structures in the body of the system. These 
structures are local realizations of the new phase \cite{nota1}.
If we further lower the
temperature, the size of the ordered regions increases, until 
most of them stick to each other and form a domain
which spans the whole system. When this happens, the system is
in a new state of order, i.e. in a new phase.

This mechanism led already at the end of the 40's \cite{onsager} to the idea that
the physics of the phase transition is basically governed by the
ordered domains built by the spin-spin correlations, and not by the individual
spins. 
If the degrees of freedom relevant for the phase change
are the ones of sets of particles, and not of
single particles,
it is likely that
they do not depend on the details of the microscopic interaction, but only
on its gross features (e.g. symmetries). 
That could explain
why whole classes of systems, ruled by
dynamics which appear very different from each other, 
happen to have the same behaviour at the phase transition.
In particular, for second-order phase transitions, 
this simple picture could allow to explain the universality of the critical
indices. 

Percolation theory \cite{stauffer} is the ideal framework 
for a geometrical description of phase transitions. The percolation 
phenomenon takes place when geometrical {\it clusters}, formed by 
elementary objects of some system, stick to each other giving rise to
an infinite network, that spans the whole system. The analogy with the 
phase transition mechanism described above is evident, but there is much more
than that. 
The real amazing thing is the fact that, 
in spite of the apparently
different nature of percolation, which is a purely geometrical 
phenomenon, and second-order thermal phase transitions, these two types of processes
are formally identical, as they are characterized by the same basic features,
which are:

\begin{itemize}
\item{Power law behaviour of the variables near the critical point, with
    relative exponents;}
\item{Scaling relations between the exponents;} 
\item{Universality of the critical indices.}
\end{itemize}

The temptation to identify a continuous thermal phase transition with a simple
percolation transition is then very strong. 
This identification is possible provided one can 
establish a one-to-one correspondence between thermal and percolation variables.
The main percolation 
variables are:

\begin{itemize}
\item{the percolation strength $P$, i.e. the probability that a site chosen
at random belongs to a percolating cluster;}
\item{the average cluster size $S$,
\begin{equation}\label{defS}
  S\,=\,\frac{\sum_{s} {{n_{s}s^2}}}{\sum_{s}{n_{s}s}}~,
\end{equation}
where $n_s$ is the number of clusters with $s$ sites and the sums exclude eventual
percolating clusters.
}
\end{itemize}

Suppose we have defined how to
group the spins of the configurations of a given model in clusters.
Such clusters are the `physical' clusters or {\it droplets} of the
model if the following conditions are satisfied:

\begin{itemize}
\item{the percolation point coincides with the thermal critical point;}
\item{the connectedness length (average cluster radius) diverges as the 
thermal correlation length (same exponent);}
\item{the percolation strength $P$ near the threshold
varies like the order parameter $m$
of the model (same exponent);}
\item{the average cluster size $S$ diverges as the 
physical susceptibility $\chi$ (same exponent).}
\end{itemize}

The first studies in this direction concerned the Ising model.
The simplest clusters one can think of   
are just the magnetic domains, i.e. the clusters built by
joining nearest-neighbour spins of the same sign. In two dimensions
these clusters indeed percolate at the critical temperature of the
Ising model \cite{connap}; however, the critical percolation exponents
differ from the thermal ones \cite{sykes}.
So, the magnetic domains are not the Ising droplets. It was soon 
realized that the problem of these clusters is that they are too big 
due to purely geometrical effects. The average cluster size would be 
non-negligible also at extremely high temperatures, just because 
spins of the same sign may happen to
lie close to each other, even in the absence of a physical correlation
between them.
In order to eliminate these artificial geometrical correlations,
one can introduce a bond probability $p_B$ and join
nearest-neighbouring spins with this probability, 
which automatically reduces the size
of the clusters.
Since the correlation
changes with the temperature $T$, $p_B$ must be 
as well a function of $T$. If one chooses 
the expression $p_B=1-\exp(-2J/kT)$ \cite{nota2} ($J$ is the Ising coupling), the
corresponding site-bond clusters are indeed the critical droplets
of the Ising model, in any dimension \cite{FK,CK}.

Nobody can yet say whether the phase transition of 
every model can be geometrically described as
a percolation transition, i.e. whether one can always define the droplets.
The Ising result, which is valid more in general for the $q$-state Potts model, 
can easily be extended to several spin systems, both discrete
and continuous \cite{san1,san2,san3}. In general, one finds that 
each interaction between a pair of spins corresponds to a bond 
in the percolation picture with an
analogous bond probability as in Ising. In models with several spin-spin interactions
of the same type (all ferro- or antiferromagnetic), one can 
still define a percolation picture \cite{san2} 
by putting bonds between any pair of interacting spins with some probability,
but the picture becomes quite involved: if two interacting spins are
far from each other, the geometrical bond
between them looks virtual, as the two spins are geometrically
disconnected. 

Recent investigations \cite{san4,san5} aimed at recovering the importance
of the role of geometrical connectivity in the mapping between percolation and 
thermal critical behaviour. In \cite{san5} it was shown that,
for a wide class of bidimensional models, one can define 
simple site-bond clusters 
which show all features the droplets
should have. 
For the models where
a rigorous mapping between percolation and critical behaviour is possible, 
such site-bond clusters are in general different from 
the "exact" droplets (see \cite{san1,san2,san3}), which are in general
more complex, even if
their behaviour at criticality is identical.
Moreover, the result remains valid as well for models
with competitive interactions (e.g. ferromagnetic + antiferromagnetic), 
for which an exact definition
of the droplets is, at present, missing.
In \cite{san5} one examined theories with center symmetry 
$Z(2)$ and $Z(3)$, such that 
their critical behaviour is in the universality class of 
the model obtained by removing all interactions except the
nearest-neighbour one (Ising for $Z(2)$, 3-state Potts
for $Z(3)$). Therefore, for these models
the nearest-neighbour spin-spin coupling is the fundamental interaction
which determines the behaviour at the phase transition. 
This is probably the reason why, if one just 
considers geometrical connections between nearest-neighbours,
weighted by some suitable bond probability, the corresponding clusters
are at least a good approximation of the critical droplets of the model.
The previous argument is of course independent of the
number $d$ of space dimensions of the system.
For this reason we believe that the result of \cite{san5}
must be valid in general, i.e. for $d>2$ as well. In this paper we review
the 2-dimensional results presented in \cite{san5} and provide 
numerical evidence, based on Monte Carlo simulations, 
that our droplet definition holds true also in three dimensions. 

We stress that the original target of
our investigations was to provide a geometrical description
in terms of percolation 
of the confinement-deconfinement transition in $SU(N)$ gauge theories.
Early attempts focused on $SU(2)$ pure gauge theory, whose deconfining
transition is second-order and in the universality class of the Ising model
\cite{engels}.
The strategy we followed at that stage was to 
approximate the gauge model by means of 
simpler Polyakov loop effective theories for which an exact droplet definition
exists \cite{san6,san7}. In this way one 
finds only
an approximate solution of the problem,
which strongly depends on the specific lattice regularization
one chooses \cite{nota3}. This seemed to us unsatisfactory: the droplet
prescription we propose here solves the problem in a simple and general way.

The paper is divided as follows: 
in Section II we discuss more in detail 
our droplet definition; in Section III 
we present the results of our simulations, 
distinguishing between spin models and
$SU(2)$ pure gauge theory; finally  
the conclusions of our work are exposed.

\section{The Droplet Definition} 

Our droplet candidates are clusters built by joining nearest-neighbour
spins of the same sign with a bond probability $p_B$: they are then uniquely 
defined once we specify $p_B$. In \cite{san5}, the following criterion
was proposed: $p_B$ is the minimal probability that still makes 
percolation possible at the critical temperature $T_c$. 
This special minimal probability, that we
will call $p_{CK}$, depends on the model at study.
We recall that  
$p_{CK}$ is in general a function of the temperature, like
in the Ising model (where $p_{CK}(T)=1-\exp(-2J/kT)$). We are only 
interested in the behaviour at the transition, i.e.
near $T_c$. 
Therefore 
what matters is basically only the value $p_{CK}(T_c)$ of  
the bond weight at $T_c$ \cite{nota3b}. This is why, referring to $p_B$ ($p_{CK}$), 
we use the term
"value" instead of "expression". 

In two dimensions, 
for any $p_B<p_{CK}$, the percolation
temperature $T_p<T_c$ and the exponents are in the universality class 
of 2D pure random percolation. On the other hand, for any $p_B>p_{CK}$
(including
the pure-site case $p_B=1$),
$T_p=T_c$, but the exponents
do not coincide either with the thermal or with the random percolation ones.
For $p_B=p_{CK}$ and only in this case, the site-bond clusters satisfy all conditions 
required for the droplets, i.e. both the critical temperature and the exponents 
of the geometrical transition are equal to the thermal counterparts.
We see that there is a whole range of $p_B$ values, i.e.
$p_{CK}\,\leq\,p_B\,\leq\,1$, for which the relative site-bond
clusters begin to percolate exactly at the onset of the thermal transition.
Such feature is specific of bidimensional lattices. In three dimensions
we shall see that there is just a single value of the 
bond probability $p_B$ for which the percolation threshold is exactly at $T_c$. 
If $p_B$ is greater than this value, the clusters begin
to percolate at some $T_p>T_c$.
However, this asymmetry between the 2D and the 3D
cases does not represent a serious problem. In the Ising model,
by using the Fortuin-Kasteleyn  
bond weight $p_{CK}=1-\exp(-2J/kT)$,
one obtains the correct droplets in any dimension \cite{FK,CK}.
In particular, for the 3-dimensional Ising model, 
$p_{CK}(T_c)$ is necessarily the unique $p_B$ value
for which the two thresholds coincide \cite{nota4}.
This special $p_B$ value is again the minimal
probability one needs in order 
to have a percolating cluster at $T_c$, so that
the "criterion of the minimal 
bond probability" adopted in \cite{san5}
would lead to the correct droplet definition
in three dimensions too, at least in the Ising case.
We then assume that such criterion is valid more in general,
and we shall verify its validity by computer simulations
of two models in three dimensions, the $O(2)$ spin model and 
$SU(2)$ lattice gauge theory.

\section{Results}

\subsection{Numerical Analysis}

Our aim is to investigate the percolation transition
of special site-bond clusters, determining in particular
the percolation temperature and the critical exponents. 

To produce the equilibrium configurations we made use of standard
Monte Carlo algorithms, like Metropolis or heat bath; for some models we
adopted cluster updates, like the Wolff algorithm,
which allows to reduce sensibly the correlation
of the data and save a lot of CPU time.
At each iteration, once 
the configuration to be examined is determined, all lattice sites 
are grouped in clusters by means of the algorithm 
devised by Hoshen and Kopelman \cite{kopelman}; for the cluster labeling we
have always used free boundary conditions.
After that we are left with a set of clusters of various sizes,
and we can calculate the percolation variables.
If a cluster connects the top with the bottom side (face in 3D)
of the lattice, we say that it percolates \cite{nota5}.
Besides the percolation strength $P$ and the average cluster size $S$, we 
also calculate the size $S_M$ of the largest cluster of the configuration, since 
from it one can derive the fractal dimension $D$ of 
the spanning cluster at the threshold \cite{nota6}.
At each iteration the energy density $\epsilon$ and the lattice average $m$ of the 
order parameter of the thermal transition
were also stored \cite{nota7}.

The first step is of course the determination of the percolation
temperature. This can be effectively done by using a variable
that can be extracted from the data sample of the 
percolation strength $P$. Suppose we have 
performed a number $N_I$ of iterations for one 
of our models at a given temperature and lattice size.
Looking at the column of the $P$ data, say
$N_P$ the number of configurations of our sample with (at least)
a percolating cluster (for those configurations
$P{\neq}0$). The ratio $N_P/N_I$ is the {\it percolation cumulant} $\Pi$,
which shares the same properties of the well-known Binder cumulant, namely:

\begin{enumerate}
\item{if one plots $\Pi$ as a function of $T$, all curves corresponding to
different lattice sizes cross at the same temperature $T_p$, which marks the
threshold of the percolation transition;}
\item{the percolation cumulants for different values of the lattice size
$L$ coincide, if considered as functions of
$t_pL^{1/\nu_p}$ ($t_p=(T-T_p)/T_p$, $\nu_p$ is the exponent of the 
connectedness length);}
\item{the value of $\Pi$ at $T_p$ is a universal quantity,
    i.e. it labels a well defined set of critical indices}.
\end{enumerate}

The first property suggests 
that it is enough to make simulations
on two different lattices to determine the
critical point. The result is of course the more precise the
larger the size of the lattices. 

After evaluating the
percolation temperature $T_p$, we extracted the values of the 
critical indices by means of standard finite-size scaling techniques at
the critical point.
If corrections to scaling
do not play an important role, 
the finite-size scaling laws of the percolation variables
at $T_p$ are given by the simple formulas

\begin{eqnarray}
P(T_p)\,&\propto&\,L^{-\beta_p/\nu_p}\\
S(T_p)\,&\propto&\,L^{\gamma_p/\nu_p}\\
S_M(T_p)\,&\propto&\,L^{D},
\end{eqnarray}

where $L$ is the lattice side and $\beta_p$, $\gamma_p$ are 
the exponents that rule the power law behaviour at criticality of
$P$ and $S$, respectively. In order to improve the 
precision of the fits and to keep 
disturbing finite-size effects under control, for each model 
four to six different lattices were used.
In all our analyses we found that corrections to scaling 
do not perturb appreciably the leading behaviour expressed 
by Eqs. (2)-(4).

\subsection{Spin Models}

We start by reviewing the investigations of the bidimensional
models presented in \cite{san5}. There, we analyzed 
two classes of systems: 
models with $Z(2)$ global symmetry and a magnetization transition
with Ising exponents and models with $Z(3)$ global symmetry  
and a magnetization transition with exponents belonging to the
2-dimensional 3-state Potts model universality class. The spin systems belonging
to the first group are:

\begin{enumerate}
\item{the Ising model, ${\cal{H}}=-J\sum_{ij}s_is_j$ \,\,\,($J>0$, $s_i={\pm}1$);}
\item{a model with nearest-neighbour (NN) ferromagnetic coupling and 
a weaker next-to-nearest (NTN) antiferromagnetic coupling: 
${\cal{H}}=-J_1\sum_{NN}s_is_j-J_2\sum_{NTN}s_is_j$ ($J_1>0$,
$J_2<0$, $|J_2/J_1|=1/10$, $s_i={\pm}1$);}
\item{the continuous Ising model, ${\cal{H}}=-J\sum_{ij}S_iS_j$ 
\,\,\,($J>0$, $-1{\leq}S_i{\leq}+1$).}
\end{enumerate}

The models belonging to the second group are:

\begin{enumerate}
\item{the 3-state Potts model, ${\cal{H}}=-J\sum_{ij}\delta(s_i,s_j)$ \,\,\,($J>0$, $s_i=1,2,3$);}
\item{a model obtained by adding to 1) 
a weaker next-to-nearest (NTN) antiferromagnetic coupling: 
${\cal{H}}=-J_1\sum_{NN}\delta(s_i,s_j)-J_2\sum_{NTN}\delta(s_i,s_j)$ \,\,\,($J_1>0$,
$J_2<0$, $|J_2/J_1|=1/10$, $s_i=1,2,3$);}
\end{enumerate}

The strategy we followed in the numerical analysis was to 
tune by hand the value of the bond probability $p_B$ until,
at the critical temperature $T_c$ of the model, the percolation cumulant
$\Pi$ takes the same value for each of the lattices we used.
The smallest $p_B$ value for which this is still possible is the 
minimal bond probability $p_{CK}$ we look for.
The threshold value of $\Pi$ gives a strong indication
on the universality class of the geometrical transition (property 3
of the percolation cumulant).
To calculate the error on $p_{CK}$ 
we decreased $p_B$ until the
$\Pi$ values of all lattices at $T_c$ 
were offset by more than one $\sigma$. 

From a strictly numerical point of view one should take care
to interpret the data of the simulations when $p_B$ is 
close to $p_{CK}$. In this case, in fact, the system
finds itself in the neighbourhood of a discontinuity and if the lattice is not
large enough, its behaviour is influenced by that.
The simulations on small lattices would produce 
configurations which represent a sort of mixture of the two situations
at $p_B=p_{CK}$ and $p_B{\neq}p_{CK}$. To recover the real behaviour
of the system one should then go to very large lattices and disregard
the small ones. 

In two dimensions, as we said above, there is a whole range of 
$p_B$ values such that the onset of the percolation transition
is exactly at the thermal critical point. The calculation of the 
exponents is then crucial to distinguish the various geometrical transitions.

\begin{figure}[htb]
  \begin{center}
    \epsfig{file=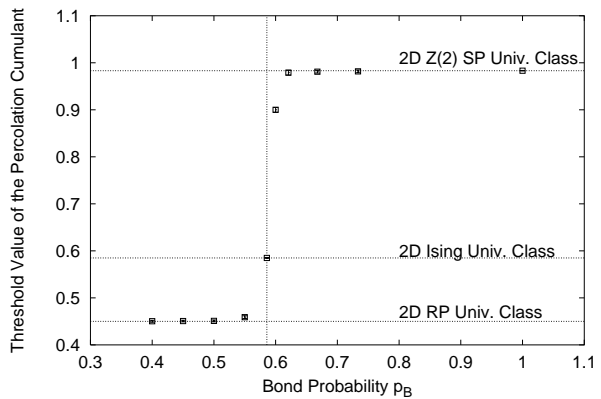,width=9cm}
    \caption{\label{fig1}{2D Ising model: variation of $\Pi$ at 
the percolation temperature $T_p$ with the bond weight $p_B$.}}
  \end{center}
\end{figure}

We start by discussing the spin systems with $Z(2)$ symmetry. 
Fig. \ref{fig1} shows the threshold value of the percolation cumulant $\Pi$
in the Ising model as a function of the bond probability $p_B$. 
For $p_B\,\approx\,p_{CK}$ we plotted the value of $\Pi$ for the
largest lattice we took ($1000^2$), as $\Pi$ changes sensibly 
with the lattice size for the reason we explained above \cite{notabis}.

The vertical dashed line in the plot marks the minimal probability
$p_{CK}=1-\exp(-2J/kT_c)=0.58578$. We see that the cumulant is quite 
stable to the right and to the left of $p_{CK}$, and that the
plateau values correspond to two different universality classes. 
For $p_B<p_{CK}$ the site-bond clusters percolate at a temperature
$T_p<T_c$, and the percolation exponents are in the 
random percolation universality class \cite{nota8}.
When $p_B>p_{CK}$ $T_p=T_c$ and the exponents belong to a special  
universality class (we call it 2D $Z(2)$ site percolation universality
class because it includes the pure site percolation
case, $p_B=1$, and is the same for all $Z(2)$ models we studied). 
For $p_B=p_{CK}$ we recover the Fortuin-Kasteleyn 
mapping \cite{FK} and the site-bond clusters are the 
exact critical droplets
of the system. We see that in this case the $\Pi$ threshold value
does not lie on either of the plateaus, since the exponents are 
now in a different universality class, i.e. in the class of the
2D Ising model. 

For the other $Z(2)$ spin systems the situation is  
analogous and we could show identical pictures as Fig. \ref{fig1},
except that the minimal probabilities $p_{CK}$ are different
($p_{CK}=0.583(1)$ for Model 2 and $p_{CK}=0.6115(9)$ for Model 3).
We remark that
for the continuous Ising model we bound nearest-neighbour
spins of the same sign, independently of their absolute values,
although they play a key role in the definition
of the "exact" droplets \cite{san1}.
However, there is no proof of the existence 
of an exact correspondence between the percolation transition
of the site-bond clusters for $p_B=p_{CK}$ and the  
magnetization transition. So it is essential
to calculate precisely the critical exponents to show
that the "minimal" clusters are indeed droplets
for the system. The results are shown in Table \ref{tab1}, where we 
can see that the agreement with the critical indices of the 2D Ising
droplets is very good. 

\begin{table}[h]
\begin{center}
\begin{tabular}{|c|c|c|c|c|}
\hline$\vphantom{\displaystyle\frac{1}{1}}$
&$\beta_p/\nu_p$ &$\gamma_p/\nu_p$  & D & $\Pi$ at $T_p$\\
\hline$\vphantom{\displaystyle\frac{1}{1}}$
2D Ising & 1/8=0.125 & 7/4=1.75&15/8=1.875 &0.585(1)\\
\hline$\vphantom{\displaystyle\frac{1}{1}}$
Model 2 & 0.131(10) & 1.742(12)&1.862(20)&0.583(4)\\
\hline$\vphantom{\displaystyle\frac{1}{1}}$
Model 3 & 0.121(9) & 1.764(14)&1.870(11)&0.587(3)\\
\hline
\end{tabular}
\caption{\label{tab1} Critical percolation indices for the site-bond 
clusters of the $Z(2)$ models 
when $p_B=p_{CK}$, compared with the values of the
2D Ising droplets.}
\end{center}
\end{table}

\begin{figure}[htb]
  \begin{center}
    \epsfig{file=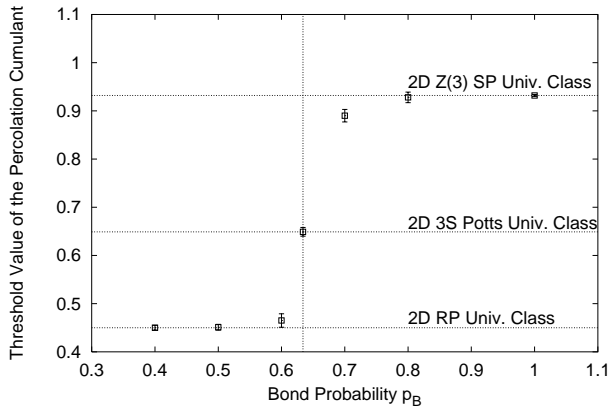,width=9cm}
    \caption{\label{fig2}{2D 3-state Potts model: variation of $\Pi$ at 
the percolation temperature $T_p$ with the bond weight $p_B$.}}
  \end{center}
\end{figure}

As far as the analysis of the $Z(3)$ spin systems is concerned,
we attain the same conclusions. Fig. \ref{fig2} shows the 
dependence on $p_B$ of $\Pi$ at the percolation threshold for the 3-state Potts model.
We notice that we obtain the same pattern we found for the 
$Z(2)$ models (see Fig. \ref{fig1}). 
Since the magnetization transition 
is characterized by exponents which are different from the Ising ones,
one expects to find another set of critical indices
to the right of $p_{CK}=0.6339736...$. As we can see from the figure,
the plateau for $p_B>p_{CK}$
lies indeed at a different height compared to the one we have in the Ising 
plot (0.932 instead of 0.9832). Analogously as we
did above, we call this new set of critical indices
2D $Z(3)$ site percolation universality class. 
The values
of these indices and of their $Z(2)$ counterparts 
were predicted in \cite{stella,vander}: our numerical
findings confirm such theoretical predictions, which 
are listed in Table \ref{tab2}.

\begin{table}[t]
\begin{center}
\begin{tabular}{|c|c|c|c|c|c|}
\hline$\vphantom{\displaystyle\frac{1}{1}}$
&$\beta_p$ &$\gamma_p$  & $\nu_p$ & D & $\Pi$ at $T_p$\\
\hline$\vphantom{\displaystyle\frac{1}{1}}$
2D Z(2) SP & 5/96 & 91/48 & 1 & 187/96 &0.9832(4)\\
\hline$\vphantom{\displaystyle\frac{1}{1}}$
2D Z(3) SP &  7/96 & 73/48 & 5/6 & 153/80 &0.932(2)\\
\hline
\end{tabular}
\caption{\label{tab2} Critical indices of the percolation transition
of site-bond clusters when $p_B>p_{CK}$, for the two 
groups of spin systems we considered.}
\end{center}
\end{table}

For the $Z(3)$ model with competitive interactions the 
results are the same, apart from the value of the 
minimal probability ($p_{CK}=0.61(1)$). Again, in order to
prove that the minimal clusters are droplets for the system we can only rely on 
the numerical evaluation of the exponents. We report our estimates in
Table \ref{tab3}: the agreement with the indices of the 
3-state Potts droplets is good.

\begin{table}[h]
\begin{center}
\begin{tabular}{|c|c|c|c|c|}
\hline$\vphantom{\displaystyle\frac{1}{1}}$
&$\beta_p/\nu_p$ &$\gamma_p/\nu_p$  & D & $\Pi$ at $T_p$\\
\hline
\hline$\vphantom{\displaystyle\frac{1}{1}}$
2D 3S Potts & 2/15& 26/15&28/15&0.649(9)\\
\hline$\vphantom{\displaystyle\frac{1}{1}}$
Model 2 & 0.143(17) & 1.725(21)&1.858(18)&0.646(11)\\
\hline
\end{tabular}
\caption{\label{tab3} Critical percolation indices for the site-bond 
clusters of the $Z(3)$ models
when $p_B=p_{CK}$.}
\end{center}
\end{table}

\begin{figure}[htb]
  \begin{center}
    \epsfig{file=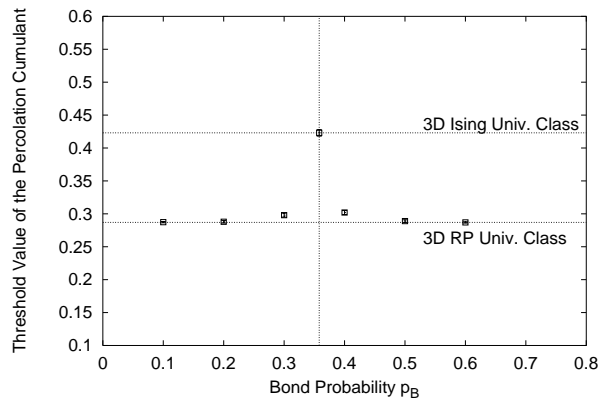,width=9cm}
    \caption{\label{fig3}{3D Ising model: variation of $\Pi$ at 
the percolation temperature $T_p$ with the bond weight $p_B$.}}
  \end{center}
\end{figure}

In three dimensions, as we said in Section II, the situation looks
quite different. We performed simulations of the 3D Ising model, using
several values for the bond probability above and below
$p_{CK}=1-\exp(-2J/kT_c)=0.35808..$. We studied the variation
of $\Pi$ with $p_B$ as we did in two dimensions: the result is shown
in Fig. \ref{fig3}. For $p_B\,\neq\,p_{CK}$,
the percolation temperature $T_p$ is different
from the magnetization temperature $T_c$ and 
we recover the 3D random percolation exponents (see \cite{nota8}).
Only for $p_B=p_{CK}$ is $T_p=T_c$
and we could eventually get the thermal critical indices
(for the 3D Ising model this is exactly what happens).

We notice that, even if the 3D pattern is very different from
the 2D one, the special (Fortuin-Kasteleyn) probability $p_{CK}$ 
is still the smallest probability for which the site-bond clusters
can form a percolating cluster at $T_c$ (for $p_B<p_{CK}$, $T_p<T_c$ and all
clusters at $T_c$ are finite). 

We want to verify whether this is also valid for other 3D spin systems,
and we analyze here the $O(2)$, or $XY$, model. For $O(n)$ models 
a rigorous mapping between percolation and critical behaviour
was established in \cite{san3}. The droplets are built in two steps:

\begin{enumerate}
\item{choose a random vector {\bf r} of $O(n)$;}
\item{bind
together any pair of nearest-neighbouring spins ${\bf s_i}$, ${\bf s_j}$ with
the probability
 
\begin{equation}
p(i,j)=1-\exp\{min[0,-2{\beta}({\bf s_i}{\cdot}{\bf r})
({\bf s_j}{\cdot}{\bf r})]\}
\end{equation}

($\beta=J/kT$).}
\end{enumerate}

Such droplets are just 
the clusters devised by Wolff in his famous algorithm \cite{wolff} for 
$O(n)$ spin systems.
We see that only pairs of 
spin vectors having both a positive/negative projection
on the random vector {\bf r} can be joined to each other. 
The random vector {\bf r}, therefore, divides
the spin space in two hemispheres, separating the spins
which have a positive projection onto it from the ones which have a negative
projection. The droplets are made out of spins which all lie
either in the one or in the other hemisphere.
In this respect, we can again speak of 'up' and 'down' spins, like
for the Ising model. In addition to that, the bond probability is
local, since it explicitly depends on the spin vectors 
${\bf s_i}$ and ${\bf s_j}$, and not only on the temperature
like the Fortuin-Kasteleyn factor.

The situation is similar as in the 2D continuous Ising model we 
considered above, and we proceeded in the same way, i.e. we reduced
the O(2) configurations to Ising configurations,  
according to the sign of the projection of the spins on {\bf r}, so disregarding
the length of the projection. The bond weight $p_B$ we introduced is 
the same for each pair of nearest-neighbouring sites.

\begin{figure}[htb]
  \begin{center}
    \epsfig{file=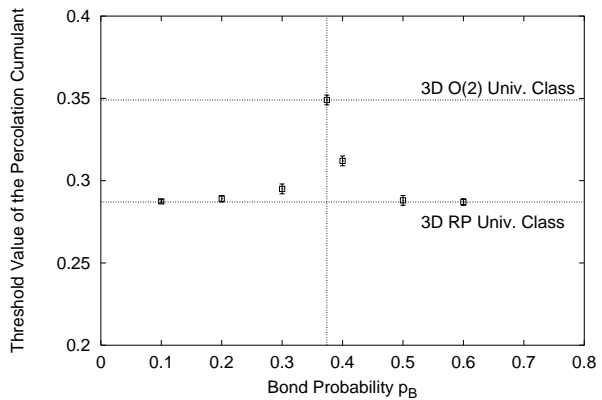,width=9cm}
    \caption{\label{fig4}{3D $O(2)$ model: variation of $\Pi$ at 
the percolation temperature $T_p$ with the bond weight $p_B$.}}
  \end{center}
\end{figure}

For our simulations we applied the Wolff algorithm and  
used four lattices: $24^3$, $48^3$, $72^3$ and $96^3$.
At each run, 40000 to 100000 measurements were taken.
We tuned the bond probability so to make the 
percolation point coincide with the magnetization point. 
At the end we found the same scenario that we had seen for the Ising
model (Fig. \ref{fig4}).

\begin{figure}[htb]
  \begin{center}
    \epsfig{file=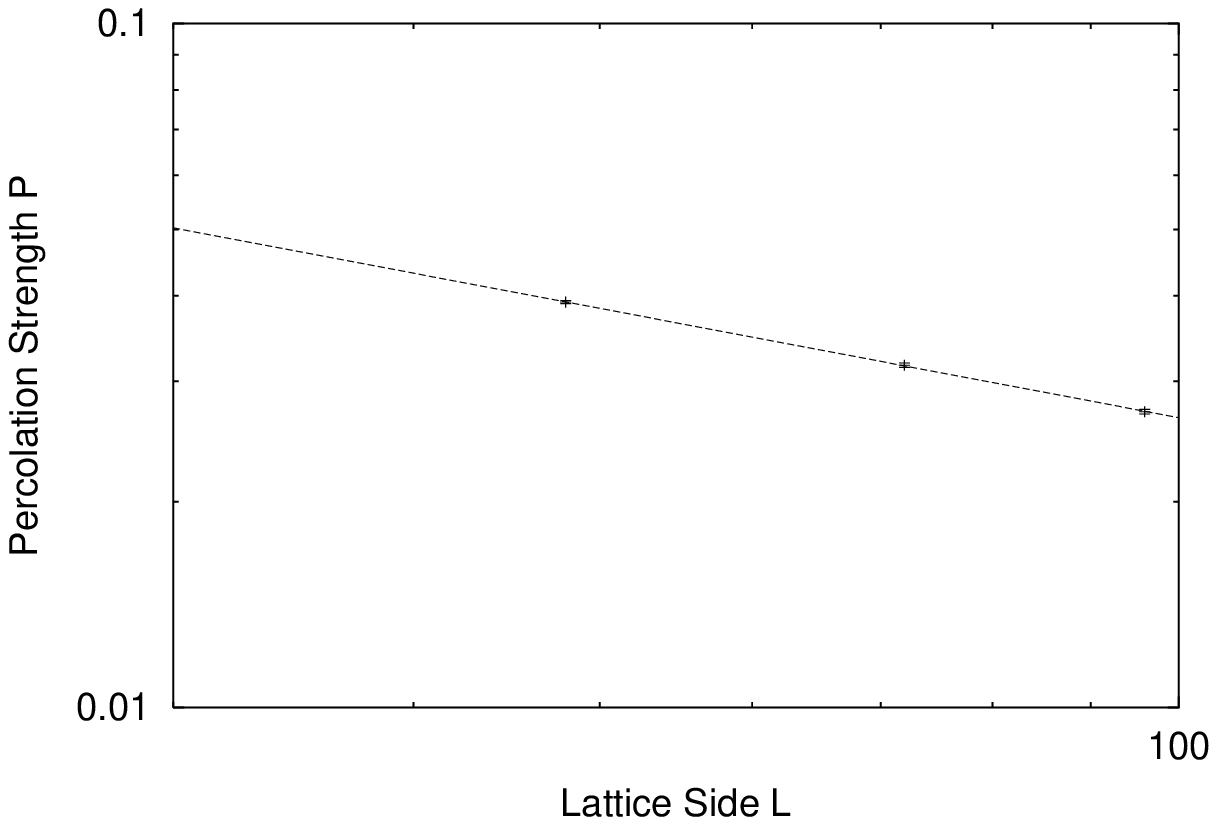,width=8cm}
    \epsfig{file=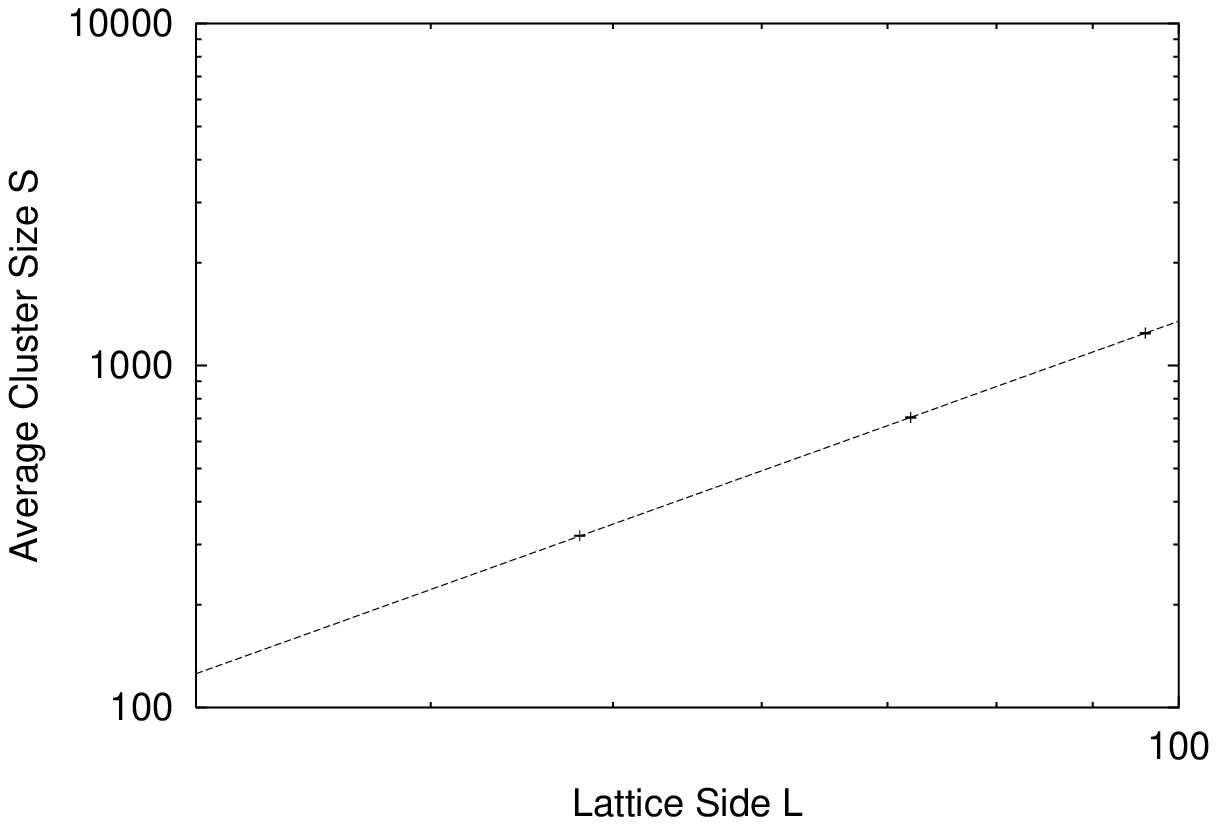,width=8cm}
    \caption{\label{fig5}{
3D $O(2)$ model: log-log plots of $P$ (top) and $S$ (bottom)
at $T_c$ versus the lattice side $L$ for the minimal site-bond clusters.}}
\end{center}
\end{figure}

Our estimate of the minimal bond probability is $p_{CK}=0.374(1)$.
Finally we calculated the critical indices of the percolation
transition when $p_B=p_{CK}$.
Fig. \ref{fig5} shows the corresponding finite size scaling  
plots of the percolation strenght $P$ (top) and the average cluster size
$S$ (bottom) at the critical point. 
The $\chi^2$ of the fits improves considerably if the smallest lattice ($24^3$)
is excluded, this is why we put only three points in the plots. 

The critical indices we extracted agree with the 
thermal $O(2)$ values (Table
\ref{tab4}).

\begin{table}[h]
\begin{center}
\begin{tabular}{|c|c|c|c|}
\hline$\vphantom{\displaystyle\frac{1}{1}}$
&$\beta_p/\nu_p$ &$\gamma_p/\nu_p$  & D \\
\hline
\hline$\vphantom{\displaystyle\frac{1}{1}}$
3D O(2) & 0.5189(3) & 1.9619(5)  & 2.4808(8)\\
\hline$\vphantom{\displaystyle\frac{1}{1}}$
Perc. Exponents & 0.530(15) & 1.971(13)& 2.484(7)\\
\hline
\end{tabular}
\caption{\label{tab4} Critical percolation indices for the site-bond 
clusters of 3D $O(2)$ when $p_B=p_{CK}$; for comparison we also report the
$O(2)$ thermal exponents (from \cite{hasen}).}
\end{center}
\end{table}

\subsection{SU(2) Pure Gauge Theory}

The deconfining transition from hadronic matter to a plasma
of quarks and gluons has been object of 
intensive investigations over the last two decades.
Though the concrete goal is to try to
produce the quark-gluon plasma by means of 
high energy heavy ion collisions in the lab,
i.e. in tiny and in general non-equilibrated 
fireballs, it is crucial from a theoretical point of view
to study the ideal situation of an infinite system of 
strongly interacting matter in thermal
equilibrium at a temperature $T$.
This could be effectively done after the discovery of the 
lattice approach \cite{wilson}, and indeed 
finite temperature Quantum
Chromodynamics ($QCD$) has been extensively simulated on the lattice
since then. 

The group
that rules 
the gauge invariance of QCD is $SU(3)$, which is
non-abelian. Because of that the gauge fields are
self-interacting and it makes sense
to study systems constitued only by gluons.
This simpler situation is described by the so-called    
$SU(3)$ pure gauge theory.  
Since any $SU(N)$ group is non-abelian, 
the study of the relative pure gauge theories
may be of interest also for $N{\neq}3$.

Suppose we have a $d$-dimensional box containing gluons at a temperature
$T$. The discretization of space-time returns a $(d+1)$-dimensional
lattice, with $N_{\sigma}$ spacings in each space direction
and $N_{\tau}$ spacings in the imaginary time (or temperature) direction. 
The partition function of finite temperature $SU(N)$ pure gauge theories
on this lattice takes the form

\begin{equation}
\label{lapart}
{\cal Z}(N_{\sigma},N_{\tau};g^2)\,=\,\int\prod\limits_{links}\,dU_{ij}\,\exp[-S(U)],
\end{equation}

where $S(U)$ is the Wilson action

\begin{equation}
\label{plaq}
S(U)\,=\,\frac{2N}{g^2}\sum\limits_{plaq}\Big(1-\frac{1}{N}Re\,Tr\,UUUU\Big).
\end{equation}

Here $g$ is the (temperature-dependent) 
coupling and $U_{ij}$ the so-called
link variable, which is a function of the gauge
fields set between a pair of nearest-neighbouring sites
$i$ and $j$. The product in Eq. (\ref{lapart})
runs over all links of the lattice, the sum in Eq. (\ref{plaq})  
over all the smallest closed paths (plaquettes),
which are formed by four links; $UUUU$ is 
the product of the link variables corresponding to each side
of a plaquette.

All $SU(N)$ pure gauge theories undergo a 
transition from a phase in which 
the gluons are bound in glueballs to a phase of free gluons. Such deconfining
transition is due to the spontaneous breaking of a global 
$Z(N)$ symmetry which results from the periodicity of the
gauge fields in the temperature direction \cite{nota9}. The order parameter is
the lattice average of the Polyakov loop, defined as

\begin{equation}
\label{pollon}
L=|{\langle}L_{\vec{x}}\rangle|
\end{equation}

with

\begin{equation}
\label{pollo}
L_{\vec{x}}\,=\,\frac{1}{N}\,Tr\,\prod\limits_{t=1}^{N_{\tau}}\,U_{\vec{x};t,t+1},
\end{equation}

The product in (\ref{pollo}) runs over all
the $U$'s in the temperature direction
taken at a given spatial site $\vec{x}$.
In the confined phase $L=0$, whereas at deconfinement
$L\,\neq\,0$. 
The main features of the deconfining transition are then all in the Polyakov
loop configurations one obtains by projecting out the temperature 
direction of the lattice through the matrix product of Eq. (\ref{pollo}).

There are conjectures suggesting that the deconfining transition of 
$SU(N)$ pure gauge theories is intimately related to 
the magnetization transition of the $N$-state
Potts model, with which they share the $Z(N)$ symmetry \cite{svet}.
In particular, it was predicted that if the 
deconfining transition is second-order, the 
critical indices are in the universality class of the corresponding
Potts model: this prediction has been confirmed
by computer simulations without exceptions.
This is actually the reason why we investigated 
simple Potts-like spin systems. As our droplet prescription
seems to work for these models, we tried to see whether 
it is correct for 
$SU(N)$ gauge theories as well, 
at least in the cases in which the deconfining
transition is continuous. 
In this way we would have
for the confinement-deconfinement transition the same geometrical picture
as for magnetization in the Potts model.
If we take a typical Polyakov loop
configuration of an $SU(N)$ theory at a certain temperature, there
will be areas where 
$L$ takes negative values, and areas where $L$ takes 
positive values. Both the positive and the negative "islands" can be 
seen as local regions of deconfinement. As long as there are finite 
islands of both signs, deconfinement remains a local phenomenon and the
whole system is in the confined phase. When one of this islands 
percolates, i.e.
it becomes infinite, we can talk of deconfinement as a global phase of the 
system.

We examined the simple case of $SU(2)$ pure gauge theory, in two and three
space dimensions. In both cases we focused on one and the same 
lattice regularization, corresponding to $N_{\tau}=2$; however, the result
should hold for any $N_{\tau}$
(see Conclusions). We stress that for $SU(2)$ the Polyakov loop
$L_{\vec{x}}$ is a real number, which can take all
possible values in a range (from -1 to +1 according to our normalization choice).
We are in the same situation as in the continuous
Ising model, and again we will consider only the sign
of $L_{\vec{x}}$, disregarding its absolute value. So, the Polyakov loop
configurations
reduce themselves to Ising-like configurations, that we treated 
in the same way as in the previous section.

\begin{figure}[htb]
  \begin{center}
    \epsfig{file=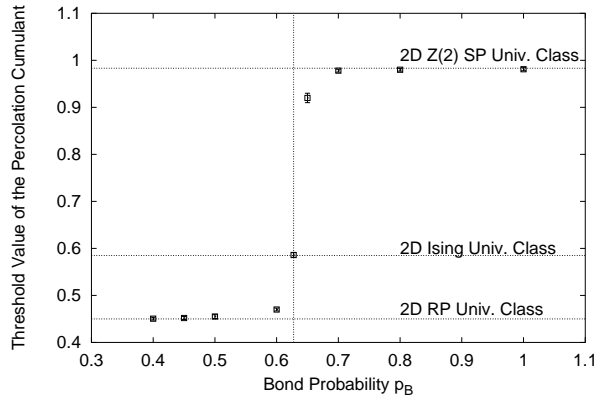,width=9cm}
    \caption{\label{fig6}{$2+1$ $SU(2)$: variation of $\Pi$ at 
the percolation temperature $T_p$ with the bond weight $p_B$.}}
  \end{center}
\end{figure}

We start by exposing the results for $2+1$ $SU(2)$. 
We carried on our simulations on four different lattices, $64^2{\times}2$,
$96^2{\times}2$, $128^2{\times}2$ and $200^2{\times}2$.
For each run we collected 
from 10000 to 40000 measurements. The result of our analysis is 
identical as for the $Z(2)$ spin systems (Fig. \ref{fig6}). 

If we compare Fig. \ref{fig6} to Fig. \ref{fig1} we see no differences,
except in the value of the 
minimal bond probability (for $SU(2)$, $p_{CK}=0.6275(5)$).

This is truly remarkable and shows the substantial analogy
of the two cases.
The exponents we calculated for $p_B=p_{CK}$ 
are in good accord with the 2D Ising ones, which shows 
that the minimal clusters are critical droplets for $SU(2)$ as well 
(Table \ref{tab5}).

\begin{table}[h]
\begin{center}
\begin{tabular}{|c|c|c|c|c|}
\hline$\vphantom{\displaystyle\frac{1}{1}}$
&$\beta_p/\nu_p$ &$\gamma_p/\nu_p$  & D & $\Pi$ at $T_p$\\
\hline$\vphantom{\displaystyle\frac{1}{1}}$
2D Ising & 1/8=0.125 & 7/4=1.75&15/8=1.875 &0.585(1)\\
\hline$\vphantom{\displaystyle\frac{1}{1}}$
$SU(2)$ Perc. & 0.140(19) & 1.761(17) & 1.882(18)& 0.586(5)\\
\hline
\end{tabular}
\caption{\label{tab5} Critical percolation indices for the site-bond 
clusters of $2+1$ $SU(2)$ ($N_{\tau}=2$)  
when $p_B=p_{CK}$, compared with the values of the
2D Ising droplets.}
\end{center}
\end{table}

\begin{figure}[htb]
  \begin{center}
    \epsfig{file=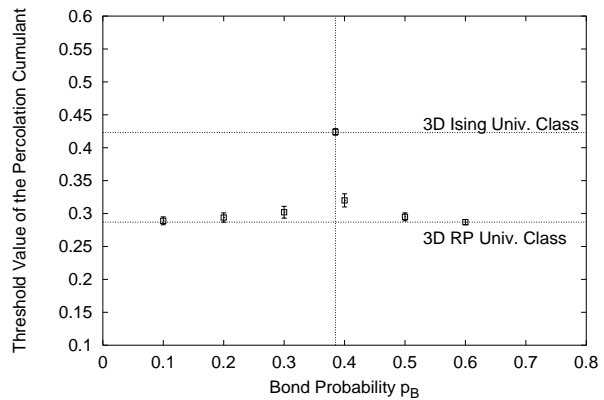,width=9cm}
    \caption{\label{fig7}{$3+1$ $SU(2)$: variation of $\Pi$ at 
the percolation temperature $T_p$ with the bond weight $p_B$.}}
  \end{center}
\end{figure}

Let us now turn to $3+1$ $SU(2)$. We used again four lattices, namely
$12^3{\times}2$, $20^3{\times}2$, $30^3{\times}2$ and $40^3{\times}2$.
The number of measurements ranged from 20000 to 50000 for each run.
The pattern we derived for $\Pi$ at the percolation threshold (Fig. \ref{fig7}) is 
the same as the one of the 3D Ising model (see Fig. \ref{fig3}), only the 
value of $p_{CK}$ is different (here is $p_{CK}=0.385(4)$). 

We notice in particular that for $p_B=p_{CK}$ the threshold value of $\Pi$
sits at the height corresponding to the 3D Ising universality class. 
The exponents relative to the transition of the minimal clusters
agree within errors with the 3D Ising exponents, as we expected (Table \ref{tab6}).

\begin{table}[h]
\begin{center}
\begin{tabular}{|c|c|c|c|c|}
\hline$\vphantom{\displaystyle\frac{1}{1}}$
&$\beta_p/\nu_p$ &$\gamma_p/\nu_p$  & D & $\Pi$ at $T_p$\\
\hline$\vphantom{\displaystyle\frac{1}{1}}$
3D Ising &0.5187(14)  & 1.963(7) &2.4817(21) &0.423(5)\\
\hline$\vphantom{\displaystyle\frac{1}{1}}$
$SU(2)$ Perc. & 0.54(3) &1.962(8)  &2.477(9) & 0.424(5)\\
\hline
\end{tabular}
\caption{\label{tab6} Critical percolation indices for the site-bond 
clusters of $3+1$ $SU(2)$ ($N_{\tau}=2$)  
when $p_B=p_{CK}$, compared with the values of the
3D Ising droplets (from \cite{isi}).}
\end{center}
\end{table}

\section{Conclusions}

We have shown that the phase transition of many 
systems can be interpreted as 
a geometrical percolation transition 
of simple site-bond clusters,
like in the Ising model.
We have found a general criterion to define the ``correct''
bond probability
$p_{CK}$ of geometrical connection between nearest-neighbouring sites
carrying spins of the same sign:
$p_{CK}$ is the minimal probability for which it is still
possible to have a percolating cluster at the critical temperature
$T_c$ of the system. We remark that this criterion looks 
a bit artificial, since it imposes by hand the coincidence of the
percolation with the thermal threshold. So, if one studies a new model,
it would be impossible to define the droplets until
one finds the critical temperature of the system. However, this is not
relevant for us, as our
aim was just to show that the droplets exist. 
We investigated a wide variety of models,
from spin systems (discrete and continuous) to 
$SU(2)$ lattice gauge theory, both in two and in three dimensions. 
To the extent of these models 
our recipe provides indeed a solution
of the problem, valid even for systems for which
``exact'' droplets could not so far be identified
(models with
competitive interactions, $SU(2)$).
The generality of the solution is clearly shown by its validity for 
a complex theory like $SU(2)$: the effective theory of the
Polyakov loop for $SU(N)$ gauge theory consists of a mixture of 
many different interactions, short- and long-ranged,
ferromagnetic and antiferromagnetic, 
including couplings between 
more than two spins (like plaquette-interactions, six-spin couplings, etc.) 
and self-interactions.
Our analysis is 
entirely numerical, but the result is most likely
exact, as it is for Ising. 

The droplet definition we propose puts in evidence the key role of 
geometrical connectivity in the mechanism of the phase transition. 
This is easy to understand for the models we have considered,
where the nearest-neighbour coupling is by far the most important
compared to eventual longer-ranged interactions and determines
the critical behaviour. However this may 
not be valid when the strength of other couplings is comparable 
to the nearest-neighbour one: in this case the phase transition
could be influenced as well by the other interactions (e.g. the critical
indices might change) and 
a droplet definition based only on nearest-neighbour 
connections is probably inadequate.

In our analysis of the models with $Z(2)$ symmetry we have seen 
that the essential spin feature for the droplet definition
is the $Z(2)$ variable, i.e. the sign of the spin. 
In fact, we built the clusters in the same way, 
no matter if the spin is discrete like in Ising or continuous like in $SU(2)$.
This shows once more 
the crucial role played by the $Z(2)$ symmetry,
whose spontaneous breaking is indeed responsible for the phase transition.
Since the relationship between $SU(N)$ pure gauge theories and
$Z(N)$ spin models holds for any $N$, 
our result should be valid for all $SU(N)$ 
pure gauge theories with a continuous deconfining transition, i.e.
also for $SU(3)$ and $SU(4)$ in $2+1$ dimensions.
The special lattice regularization of the gauge theory does not play a role,
as the critical behaviour has the same features 
for any $N_{\tau}$. So, we found that the confinement-deconfinement 
phase transition of $SU(N)$ pure gauge theory, if second-order, 
is equivalent to a percolation transition
of special site-bond clusters of like-signed Polyakov loops: this is the most important
result of our work. Very recent studies show that percolation can help to describe 
as well the chiral transition of special fermion lattice models \cite{moro}.

We focused only on systems with a continuous second-order transition because
the percolation phenomenon is typically smooth. This does not mean 
that we cannot use percolation to describe discontinuous phase changes.
The Fortuin-Kasteleyn mapping \cite{FK} is valid for any $q-$state Potts model
in any dimension, and the equivalence between the magnetization
and the percolation order parameter holds for first-order phase transitions as well.
In this case at the critical point there is a coexistence of a paramagnetic
and a ferromagnetic
phase, which correspond
to two different 
``geometrical phases'', characterized by
small and large clusters, respectively. Therefore, the percolation order parameter 
jumps at the threshold exactly as the magnetization (from zero to a non-zero value). 
Because of that we believe that the criterion of the minimal bond probability
we adopted here
can be directly extended to systems undergoing a discontinuous phase transition.
Now it is however more difficult to support the idea of droplets
because there are no critical indices to reproduce. 
One could nevertheless
try to see whether the percolation variables ($P$, $S$) vary
near $T_c$ like their thermal counterparts ($m$, $\chi$).
In this way one could study in particular the ``physical'' 
confinement-deconfinement transition of $SU(3)$ pure gauge theory 
in $3+1$ dimensions (quenched $QCD$).

\begin{acknowledgments}

I am indebted to H. Satz, A. Coniglio and Ph. Blanchard for
many helpful discussions. 
The TMR network ERBFMRX-CT-970122 and the DFG Forschergruppe
under grant FOR 339/1-2
are gratefully acknowledged.

\end{acknowledgments}


\begin{thebibliography}{99}
\bibitem{nota1} These ordered regions are, for instance, magnetic
domains in a piece of iron, or liquid drops in steam.
\bibitem{nota2} We remark that $p_B$ is always less than one except 
at zero temperature. That means that, for any $T>0$, the geometrical 
correlations of the spins are stronger than the physical correlations between them.
In particular, for $T\,\rightarrow\,\infty$, $p_B\,\rightarrow\,0$ and the
clusters consist mainly of isolated spins.
\bibitem{nota3} The studies reported in \cite{san6,san7} concerned 
lattice regularizations of $2+1$ and $3+1$ $SU(2)$ gauge theory 
which correspond to the so-called strong coupling limit: the 
number $N_{\tau}$ of lattice spacings in the temperature direction is then 
constrained to be small, e.g. $N_{\tau}=2,3,4$. The proposed percolation picture
fails for $N_{\tau}\,{\gg}\,4$.
\bibitem{nota3b} By approximating the bond weight $p_{CK}(T)$ with 
its value $p_{CK}(T_c)$, the power law behaviour of the percolation variables
at criticality would not be appreciably affected and the 
critical exponents would remain the same. Besides, we calculated
the exponents through 
finite size scaling fits of the percolation variables at $T_c$, so that
the above-mentioned approximation does not affect at all our numerical results.
\bibitem{nota4} We refer here explicitly to three dimensions because
we could prove our conjecture numerically for models in 3D. However, 
at least for the Ising model, whatever we said about the 3D case
is valid in general for any dimension higher than two.
\bibitem{nota6} Between the size $s$  
and the radius $R$ of a cluster 
there is a relation of this type: $s\,\propto\,R^D$.
The exponent $D$ is the {\it fractal dimension} of the cluster.
\bibitem{nota7} This is necessary to determine the thermal critical 
temperature of some of the models we have studied, for which 
it is unknown.
\bibitem{notabis} Very near the critical probability $p_{CK}$, 
the infinite volume limit behaviour cannot be reproduced even on the largest
lattice we have used. Therefore, the value of $\Pi$ at the threshold on the
largest lattice still differs
from the asymptotic universal value by some amount and the corresponding
point in plots like Fig. \ref{fig1} does not sit on the relative plateau.
\bibitem{nota8} If the percolation transition occurs at $T_p\,{\neq}\,T_c$,
the physical correlation length $\xi(T_p)$ is finite; therefore, 
clusters separated by a distance $d>\xi(T_p)$ are not influenced
by each other, and the long-range behaviour 
is the same as in random percolation.
\bibitem{nota9} $Z(N)$ is the {\it center} of the $SU(N)$ group, i.e. 
the subset of those elements which commute with each element
of $SU(N)$. Such elements are the matrices obtained by multiplying 
the unit matrix by the complex $N$-roots of $1$.
\end{thebibliography}
\end{document}